\documentclass{elsart}
 \usepackage{amsmath, amssymb, mathrsfs, dcolumn,mathptmx}
\usepackage{natbib, graphicx}
\begin{document}
\runauthor{Moortgat}
\begin{frontmatter}
\title{Particle-in-cell simulations of fast collisionless reconnection in GRB outflows.}
\author[pas]{Joachim Moortgat\thanksref{nsf}}
\author[lle]{Eric G. Blackman$^{\mathrm{a},}$\thanksref{nsf}}
\thanks[nsf]{Partially supported by NSF grants AST-0406799, AST-0406823 \& NASA grant ATP04-0000-0016 (NNG05GH61G).}
\author[lle]{Chuang Ren$^{\mathrm{a},\ \mathrm{b},}$\thanksref{doe}}
\thanks[doe]{Partially supported by DOE grant DE-FG02-06ER54879.}
\author[engineering]{Xianglong Kong}
\author[lle]{Rui Yan}

\address[pas]{Department of Physics and Astronomy, University of Rochester, NY, USA}
\address[engineering]{Mechanical Engineering 
    and Physics,  University of Rochester, NY, USA}
\address[lle]{Laboratory for Laser Energetics, Rochester, NY, USA}

\begin{abstract}
We present preliminary results of particle-in-cell ({\sc pic}) simulations of collisionless magnetic reconnection for conditions that apply to magnetically dominated pair plasma jets such as those in gamma-ray bursts ({\sc grb}). We find similar behaviour to previous authors but with reduced reconnection efficiency.
This results  because we include the full electromagnetic field
dynamically and allow electric field fluctuations to grow. Although weak, these fluctuations impede early $x$-point formation by periodically decelerating and even reversing 
the inflow.  
\end{abstract}
\end{frontmatter}

\section{Introduction}

Presently, few testable theoretical predictions or observational diagnostics 
exist to robustly constrain whether GRB outflows are initially dominated 
by particles or  Poynting flux, and what fraction of the particles are 
baryons vs. pair plasma \citep{spruit2, koers}. 
Whatever the case close to the engine,
at large distances observations imply that particles are accelerated to highly relativistic velocities and dominate the energy budget. 
Magnetic reconnection is a plausible mechanism of particle acceleration in GRB jet plasma if the free energy is carried by magnetic fields.

Reconnection has mainly been studied in the lab,  the solar corona,  and particularly the Earth magnetosphere, which produced the  `Geospace Environmental Modeling (GEM) reconnection challenge': 
Here a simplified reconnection problem was studied and compared 
using several MHD, hybrid and PIC codes \citep{gem}.
Much less attention has been paid --either analytical or numerical-- to the fundamental reconnection mechanisms in either the relativistic regime or for a pair plasma.

Reconnection in a (relativistic) pair plasma is very different from an electron-ion plasma.
For the latter, the difference in skin depth between the electrons and ions and the corresponding scale lengths of the current sheet and reconnection region play an important role; reconnection is facilitated by Whistler modes, which can be related to the Hall current in the 
non-ideal generalized Ohm's law. In a pair plasma, however, the contributions from the electron and positron equations of motion to the Hall current cancel and Whistler waves don't exist \citep{blackman}. The non-ideal term in Ohm's law that allows field lines to reconnect is an effective enhanced resistivity related to the electron and positron pressure tensors and the tearing mode is the dominant instability.

To improve our understanding of magnetic reconnection for conditions applicable to GRB outflows we are performing simulations with a fully explicit and highly parallelized relativistic PIC code, {\em Osiris} \citep{osiris}.
By tuning the plasma parameters from non to highly relativistic values 
and comparing results for a pure pair plasma and a plasma with an increasing proton fraction, we can obtain different particle acceleration spectra
to be used to construct observational diagnostics of the jet's nature. 
This complements the existing PIC simulation work in the GRB 
outflow context which has focused on shock acceleration via the Weibel instability in ion-electron plasmas \citep{medvedev}, or magnetic pulse acceleration without magnetic reconnection \citep{liang}.  Spectra from each of these classes of simulation
should ultimately be compared.

\section{Set-up}
\vspace{-1em}\begin{figure}[!h]
\centerline{
%\resizebox{0.4\hsize}{!}{\includegraphics{B1B2}}
\resizebox{0.3\hsize}{!}{\includegraphics{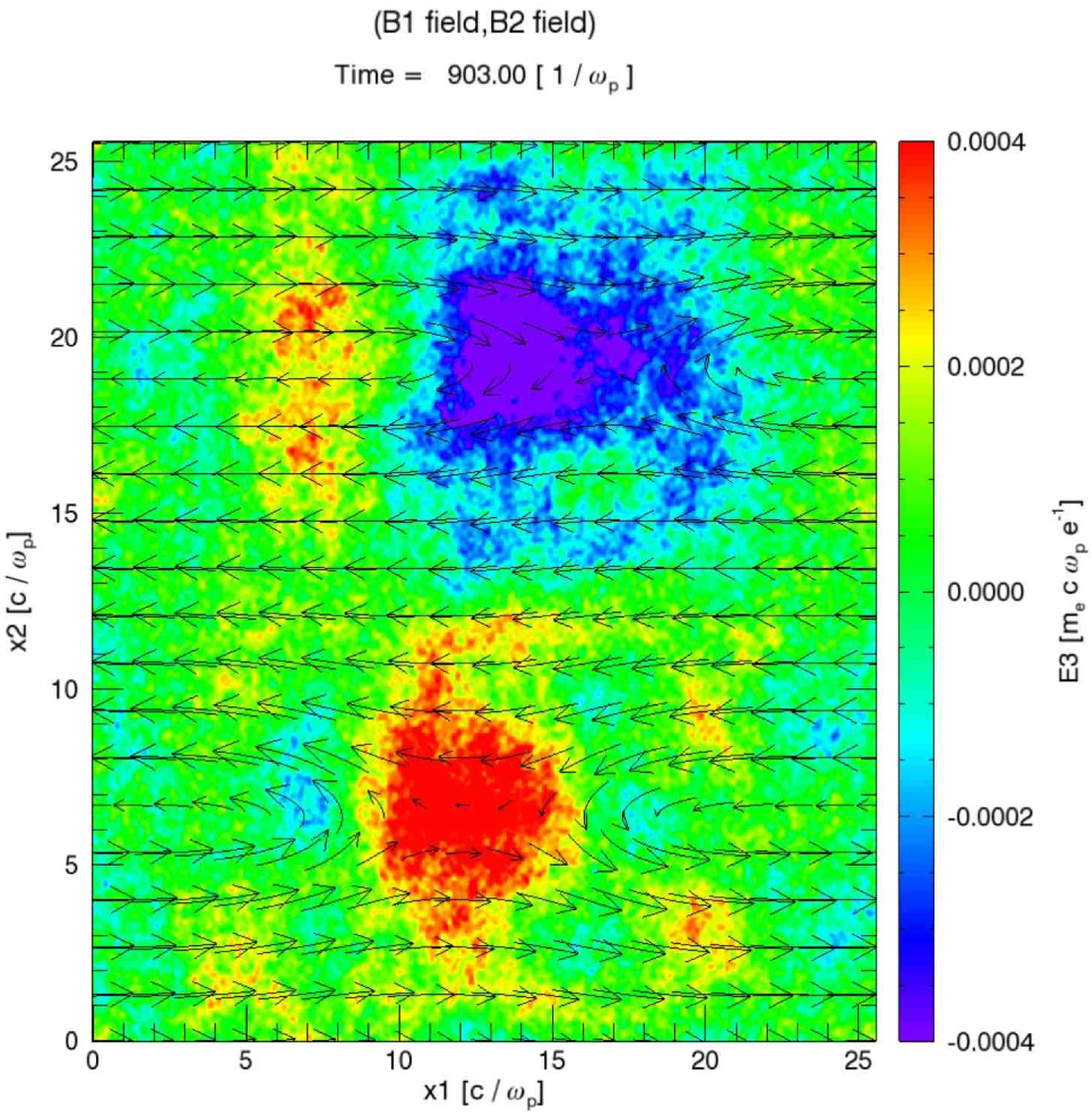}}
\resizebox{0.6\hsize}{!}{\includegraphics{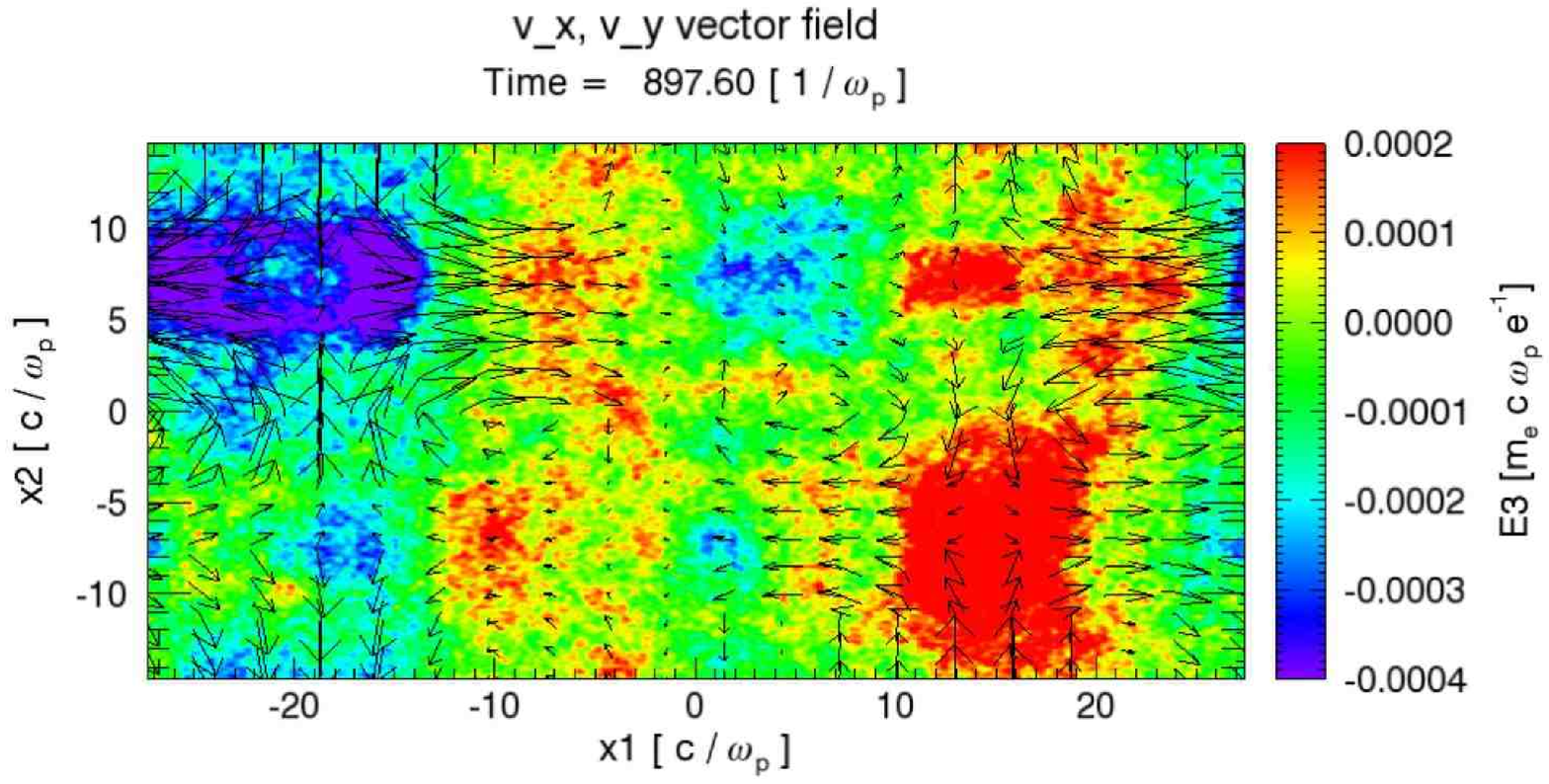}}
}
\caption{$B_{1}$-$B_{2}$ (left) and $v_{1}$-$v_{2}$ (right) vector fields during reconnection around one $x$-point on each current sheet, with $E_{3}$ as background (color), for two different simulations.}
\label{f1}
\end{figure}

A simple geometry for studying reconnection is the Harris current sheet, where two regions of nearly uniform magnetic field of opposite sign are driven into a thin current sheet by their own magnetic pressure. If the plasma can deviate from ideal MHD, the field lines can slip and reconnect at one or more $x$-points, after which the reconnected field lines are strongly bent. When those field lines straigten to minimise their tension, particles in the outflow region are accelerated.
Free magnetic energy is thus converted to particle energy in the process. 
Using the Harris configuration, we can compare to the results in the {\sc gem} reconnection challenge \citep{gem} and its extension to pair plasma\citep{bessho}. 

The simulations presented here are for a pair plasma and have two current sheets to facilitate use of periodic boundary conditions on all walls. 
In each half-plane we have an initial magnetic field normalized as:
$\vec{B} = \frac{v_{\mathrm{A}}}{c} \left[\frac{m_{e}c  \ \omega_{p}}{e}\right] \tanh(x_{2}/a)\hat{x}_{1}$, 
where $a=d_{i}/2$ is the thickness of the current sheet and $d_{i}=c/\omega_{p}$ is the skin depth in terms of the plasma frequency $\omega_{p} = \sqrt{4\pi n_{0} e^{2}/m_{e}} = c \Omega_{c}/(\sqrt{2} v_{\mathrm{A}})$ and we choose a non-relativistic) Alfv{\'e}n velocity $v_{\mathrm{A}} = B_{0}/\sqrt{8\pi m_{e}n_{0}} = 0.05c$. 
In a steady state ($\partial\vec{E}/\partial t = 0$) we have $\vec{j} = -2 e n_{0} v_{\mathrm{A}}\cosh^{-2}(x_{2}/a)$ and $
n_{e} = n_{p}= n_{0}\cosh^{-2}(x_{2}/a)$ with thermal velocity spread $(v_{\mathrm{th}})_{x, y, z} = v_\mathrm{A}/\sqrt{2}$, such that $p = 2 n_{0} k T = B_{0}/8\pi$.
A small uniform particle density $n_\mathrm{b} = 0.2 n_{0}$ is added to replenish the accelerated particles leaving the 
reconnection region and the total number of particles is $10^{8}-10^{9}$.
The spatial and time resolutions are $\Delta x = \Delta y = 0.05 d_{i}$ and $\omega_{p}\Delta t = 0.035$ with typical runs exceeding $t =10^{3}/\omega_{p}$. Most simulations use a $L_{x}\times L_{y} = 512\times 512$ grid of dimension $-12.8 d_{i} < x_{1,2} < 12.8 d_{i}$, but for some we double the length along the current-sheet to reduce the effect of  periodic boundaries. To quickly force the system into the non-linear regime, sometimes we impose an initial  perturbation, $B_{x} = 10^{-4}\cos \frac{2 \pi x_{1}}{L_{x}} \cos \frac{2 \pi x_{2}}{L_{y}}$, $B_{y} = 10^{-4} 
\sin \frac{2 \pi x_{1}}{L_{x}} \sin \frac{2 \pi x_{2}}{L_{y}}$.

\section{Results}
%\begin{figure}[!h]
%\resizebox{\hsize}{!}{\includegraphics{B1B2_E3_903_both}}
%\caption{$E_{3}$ (color); $B_{1}$-$B_{2}$ vector field.}
%\label{b1b2e3}
%\end{figure}
Figure~\ref{f1} (left) shows the magnetic field at $\Omega_{c } t = 64$ when the reconnection has reached a quasi steady state, and the reconnection electric field is well developed. This simulation was started from a perturbed equilibrium magnetic field.
The electric field at the $x$-points is $E_{3} \sim \pm 0.16 B_{0} v_{A}$, which is slightly lower than in \citet{gem} and \citet{bessho}.
When we start from numerical noise, the secondary tearing mode quickly develops, and the magnetic field fragments into a series of small  $x$- and $o$-points that move under the magnetic pressure and finally merge into one dominant $x$-point on each current-sheet. 
The right panel shows the in- and outflow velocities ($v_{1}$-$v_{2}$) around the $x$-points. One can see the particles flowing in with inflow velocity from $| E_{z} | \simeq | v_{in} B_{0} |$ and being accelerated along the $x_{1}$ axis. In this figure the effect of the periodic boundary conditions is already becoming visible with accelerated particles re-entering the box, but they're not yet influencing the reconnection region.
%\begin{figure}[!h]
%\resizebox{\hsize}{!}{\includegraphics{vxvyE3_bg_897_60}}
%\caption{$E_{3}$ (color); $v_{1}$-$v_{2}$ vector field.}
%\label{vxvyE3}
%\end{figure}

\section{Discussion}
\vspace{-1em}
In all simulations of non-relativistic reconnection we've seen
 in the literature, the fluctuating electric field is either neglected by not evolving the displacement current (common in MHD codes), or artificially eliminated at each time-step \citep{bessho}. {\em Osiris} does neither and evolves the full fields, including all electromagnetic wave modes. Doing so reveals weak 
% random electric field fluctuations in $E_{x}$ and $E_{y}$  (due to the thermal velocities of the particles) and 
plane-wave like fluctuations in $E_{3}$ (due partly to numerical noise, and partly to particles being allowed to cross the current sheet that don't cancel out exactly). When this oscillating component has the `wrong' sign, the flow of plasma into the current sheet is impeded, and initially this component is strong enough to reverse the bulk plasma motion away from the current sheet. Only during the wave-phase where $E_{3}$ has the proper sign do the particles flow into the current sheet. Eventually, the $x$-points and a mean steady reconnection electric field develop and fast non-linear reconnection is achieved. The final in- and outflow velocities reached in our simulations are somewhat slower than in previous work, and the evolution to a steady-state 
takes a bit longer.

\section{Future work} 
\vspace{-1em}
Currently, we are analysing larger runs on the NERSC clusters using more particles, higher resolution, and larger aspect ratio along the sheet. 
Furthermore, the implemented distribution functions in {\em Osiris} are only valid in the mildly relativistic regime, so we're implementing the proper relativistic J{\"u}ttner-Synge distribution. Then we can proceed
to vary the Alfv{\'e}n velocity arbitrarily, adding a small fraction of baryons, and 3D generalizations where possible.
Finally, we will calculate the radiation spectra. If the synchrotron cooling time is long compared to the reconnection rate, 
this can be done post-procees.  If the cooling time is fast, but the plasma is optically thin to the produced radiation, we must calculate the radiation self-consistently at each time-step.
Differences in the spectra produced by reconnection in a magnetically dominated wind (AC model \citet{spruit2}), a magnetic pulse model, or shock-acceleration in a matter dominated  model,  will provide a useful observational diagnostic to distinguish between the three.

\end{document}